\documentclass{iaus}
\usepackage{graphicx}

\title[Host Galaxies of GRBs] %% give here short title %%
{Host Galaxies of Gamma-Ray Bursts}

\author[Emily M. Levesque]   %% give here short author list %%
{Emily M. Levesque$^{1,2}$}

\affiliation{$^1$CASA, Department of Astrophysical and Planetary Sciences, University of Colorado 389-UCB, Boulder, CO 80309, USA $^2$Einstein Fellow \\email: {\tt Emily.Levesque@colorado.edu}}
\pubyear{2012}
\volume{279}  %% insert here IAU Symposium No.
\pagerange{xxx--yyy}
\jname{Death of Massive Stars: Supernovae and Gamma-Ray Bursts}
\editors{P. Roming, N. Kawai \& E. Pian, eds.}
\begin{document}

\maketitle

\begin{abstract}
Host galaxies are an excellent means of probing the natal environments that generate gamma-ray bursts (GRBs). Recent work on the host galaxies of short-duration GRBs has offered new insights into the parent stellar populations and ages of their enigmatic progenitors. Similarly, surveys of long-duration GRB (LGRB) host environments and their ISM properties have produced intriguing new results with important implications for long GRB progenitor models. These host studies are also critical in evaluating the utility of LGRBs as potential tracers of star formation and metallicity at high redshifts. I will summarize the latest research on LGRB host galaxies, and discuss the resulting impact on our understanding of these events' progenitors, energetics, and cosmological applications. 
\keywords{gamma rays: bursts, galaxies: abundances, galaxies: starburst}
%% add here a maximum of 10 keywords, to be taken form the file <Keywords.txt>
\end{abstract}

%\firstsection % if your document starts with a section,
              % remove some space above using this command.
\section{Introduction}
Long-duration gamma-ray bursts (LGRBs), associated with the core-collapse deaths of massive stars, are among the most energetic events observed in our universe. As a result, they are widely cited as powerful and potentially unbiased tracers of the star formation and metallicity history of the universe out to $z \sim 8$ (e.g. Bloom et al.\ 2002, Fynbo et al.\ 2007, Chary et al.\ 2007, Savaglio et al.\ 2009). However, in recent years potential biases in the star-forming galaxy population sampled by LGRBs have become a matter of debate. Recent work on a small number of nearby LGRBs suggested a connection between LGRBs and low-metallicity environments (e.g. Fruchter et al.\ 2006, Wainwright et al.\ 2007). Nearby host galaxies appeared to fall below the luminosity-metallicity and mass-metallicity relations for star-forming galaxies out to $z \sim 1$ (e.g. Modjaz et al.\ 2008, Kocevski et al. 2009, Levesque et al.\ 2010a,b). These results could potentially introduce key biases that would impact the use of LGRBs as cosmic probes. 

A metallicity bias, or some correlation between metallicity and LGRB host or explosive properties, is indeed expected under the most commonly-cited progenitor model for LGRBs, the collapsar model (Woosley 1993). Under the classical assumptions of stellar evolutionary theory, the progenitor is a single rapidly-rotating massive star which maintains a high enough angular momentum over its lifetime to generate an LGRB from core-collapse to an accreting black hole. In addition, LGRBs have been observationally associated with broad-lined Type Ic supernovae (e.g. Galama et al.\ 1998, Stanek et al.\ 2003, Malesani et al.\ 2004, Modjaz et al.\ 2006, Starling et al.\ 2011), requiring the progenitors to have shed mass, and therefore angular momentum, as a means of stripping away their outer H and He shells. Mass loss rates for these evolved massive stars are dependent on stellar winds (Vink \& de Koter 2005), which in turn are dependent on the stars' metallicity (Kudritzki 2002, Vink et al.\ 2001). For young massive stars, the metallicities of their natal environments can be adopted as the metallicities of the stars themselves. It therefore stands to reason that the wind-driven mass loss rates in high-metallicity environments would rob the stars of too much angular momentum, preventing them from rotating rapidly enough to produce a LGRB and suggesting that LGRBs should either be restricted to low-metallicity environments (e.g. Hirschi et al. 2005, Yoon et al. 2006, Woosley \& Heger 2006), or produce weaker explosions at higher metallicities (MacFadyen \& Woosley 1999). 

The work presented here originally aimed to observationally confirm and quantify the predicted role of metallicity in LGRB production and progenitor evolution. However, the results illustrate that the effects of metallicity on LGRBs are complex and do not agree with these expectations, suggesting that the predictions of stellar evolutionary theory and progenitor models may require further development.

\section{The Mass-Metallicity Relation for LGRBs}
\begin{figure}[b]
% \vspace*{-2.0 cm}
\begin{center}
 \includegraphics[width=5in]{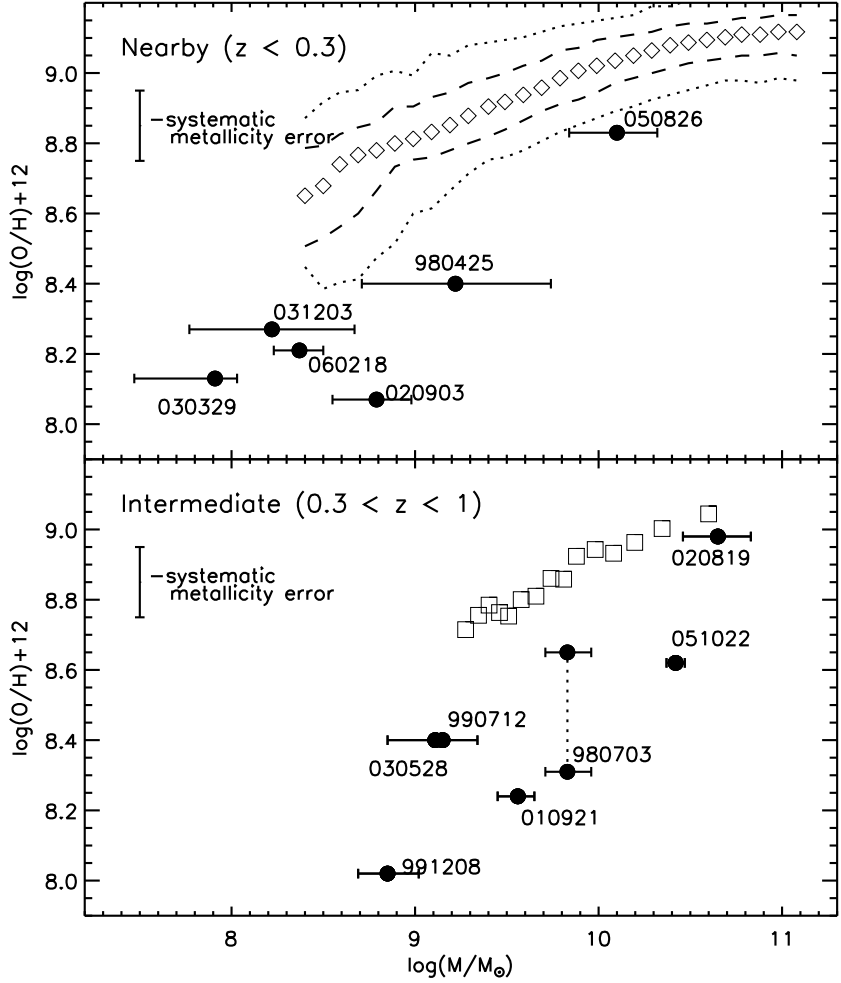} 
% \vspace*{-1.0 cm}
 \caption{Adapted from Levesque et al.\ (2010b); the mass-metallicity relation for nearby ($z < 0.3$, top) and intermediate-redshift ($0.3 < z < 1$, bottom) LGRB host galaxies (filled circles). The nearby LGRB hosts are compared to binned mass-metallicity data for a sample of $\sim$53,000 SDSS star-forming galaxies, where the open diamonds represent the median of each bin and the dashed/dotted lines show the contours that include 68\%/95\% of the data (Tremonti et al.\ 2004). For the intermediate-redshift hosts we plot binned mass-metallicity data for a sample of 1330 emission line galaxies from the DEEP2 survey (open squares; Zahid et al.\ 2011). For the $z = 0.966$ host galaxy of GRB 980703, where we cannot distinguish between the upper and lower metallicities given by the $R_{23}$ diagnostic, we plot both metallicities and connect the data point with a dotted line to indicate their common origin from the same host spectrum.}
   \label{fig1}
\end{center}
\end{figure}
In Levesque et al.\ (2010a,b) we conducted a uniform rest-frame optical spectroscopic survey of $z < 1$ LGRB host galaxies, using the Keck telescopes at Mauna Kea Observatory and the Magellan telescopes at Las Campanas Observatory. The sample was restricted to confirmed long-duration bursts with well-associated and observable host galaxies. From these spectra we were able to determine a number of key parameters for the star-forming LGRB host galaxies, including metallicity, ionization parameter, young stellar population age, SFR, and stellar mass. The primary metallicity diagnostic used in this work was the ([OIII] $\lambda$5007 + [OIII] $\lambda$4959 + [OII] $\lambda$3727)/H$\beta$ ($R_{23}$) diagnostic (Kewley \& Dopita 2002, Kobulnicky \& Kewley 2004); for our full sample we found an average $R_{23}$ metallicity of log(O/H) + 12 = 8.4 $\pm$ 0.3. Our stellar mass estimates were determined using the {\it Le Phare} code (Ilbert et al.\ 2009), fitting multi-band photometry for the host galaxies (Savaglio et al.\ 2009) to stellar population synthesis models adopting a Chabrier IMF, the Bruzual \& Charlot synthetic stellar templates, and the Calzetti extinction law (Bruzual \& Charlot 2003, Chabrier 2003, Calzetti et al.\ 2000). The fitting yielded a stellar mass probability distribution for each host galaxy, with the median of the distribution serving as our estimate of the final stellar mass. For our sample, we found a mean stellar mass of log($M_*/M_{\odot}) = 9.25^{+0.19}_{-0.23}$.

These metallicities and stellar masses were used to construct a mass-metallicity relation for LGRB host galaxies, which we plot in Figure 1. For comparison, we also compare our results to two samples of star-forming galaxies with comparable redshifts. The nearby ($z < 0.3$) LGRB hosts are compared to $\sim53,000$ star-forming SDSS galaxies, while the intermediate-redshift ($0.3 < z < 1$) hosts are compared to 1,330 galaxies from the Deep Extragalactic Evolutionary Probe 2 (DEEP2) survey (Tremonti et al.\ 2004, Zahid et al.\ 2011). Surprisingly, we found a strong and statistically significant correlation between stellar mass and metallicity for LGRB hosts out to $z < 1$ (Pearson's $r = 0.80$, $p=0.001$), with the relation showing no evidence for a clear metallicity cut-off above which LGRBs cannot be formed - instead, the overall LGRB mass-metallicity relation is offset from the mass-metallicity relation for star-forming galaxies by an average of $-0.42 \pm 0.18$ dex in metallicity. The phenomenological explanation for this offset is unclear.

\section{Energetics and Host Metallicity in LGRBs}
\begin{figure}[b]
% \vspace*{-2.0 cm}
\begin{center}
 \includegraphics[width=5in]{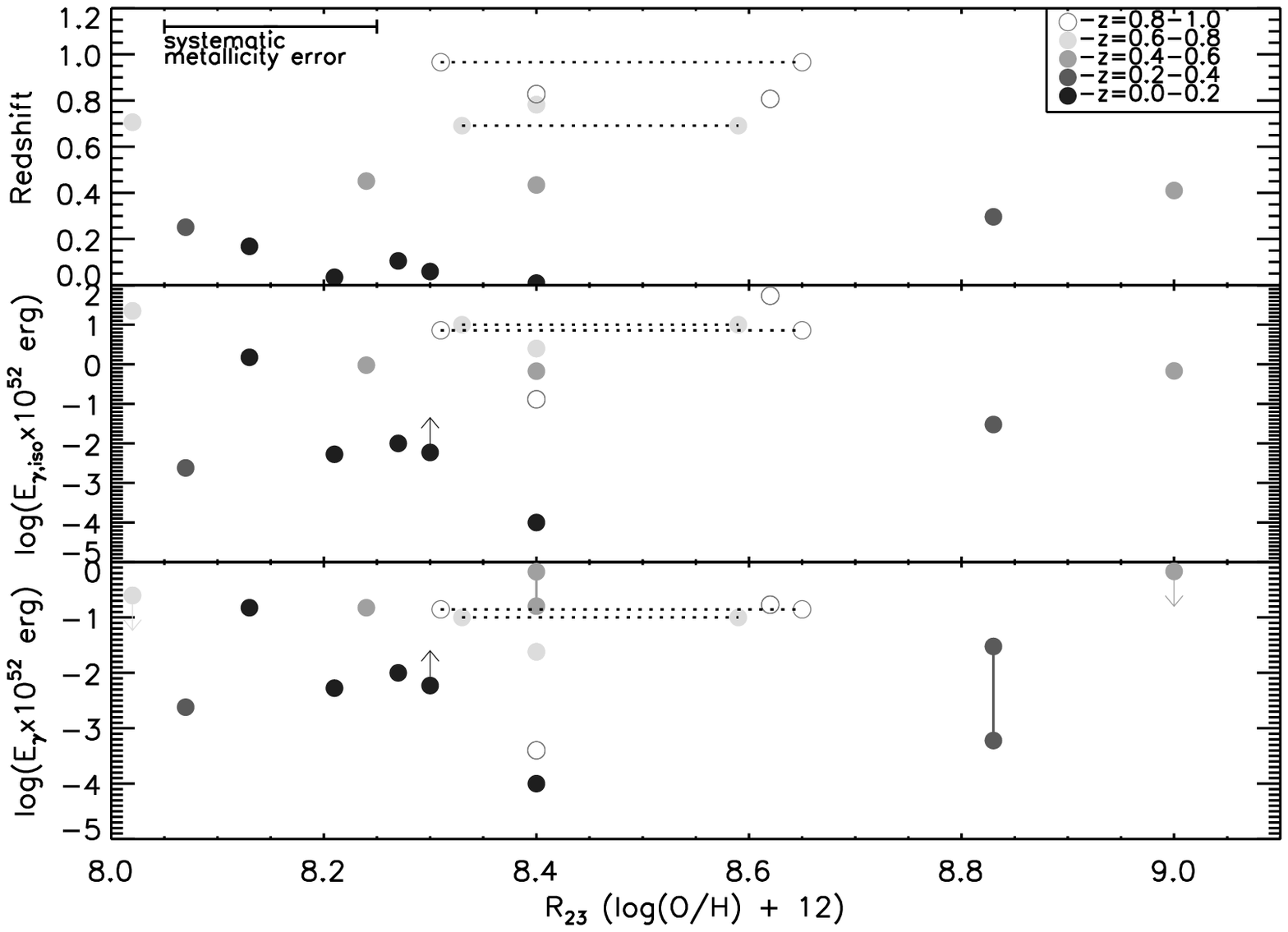} 
% \vspace*{-1.0 cm}
 \caption{Adapted from Levesque et al.\ (2010c); metallicity vs. redshift (top), $E_{\gamma, iso}$ (center), and $E_{\gamma}$ (bottom). The hosts are separated into redshift bins in order to better illustrate redshift effects. Two hosts with both lower- and upper-branch$R_{23}$ metallicities (the hosts of GRB 020405 at $z = 0.691$ and GRB 980703 at $z = 0.966$) are shown as lower and upper data points connected by dotted lines. Upper and lower limits are indicated by arrows. Hosts with both upper {\it and} lower limits on their $E_{\gamma}$ values are shown as data points connected by solid lines.}
   \label{fig1}
\end{center}
\end{figure}
Lacking observational evidence for a pure cut-off metallicity for LGRB formation, we instead consider the possibility that LGRBs at high metallicity may simply produce less energetic explosions; that is, explosions with a lower isotropic ($E_{\gamma, iso}$) or beaming-corrected ($E_{\gamma} = E_{\gamma, iso} \times 1 - cos(\theta_j)$ energy release in the gamma-ray regime, where $\theta_j$ is the GRB jet opening angle) energy release (Frail et al.\ 2001). Previous studies of several local LGRBs suggested a strong correlation between these parameters, with high-metallicity LGRBs producing markedly-lower $E_{\gamma, iso}$ (Stanek et al.\ 2006). By combining energetic parameters available in the literature with our previously-determined host galaxy metallicities, we were able to reproduce this comparison (Levesque et al.\ 2010c; Figure 2). A comparison with redshift was considered as well, to highlight any potential correlation that may appear as an artifact of metallicity evolution with redshift. However, we found that there is no statistically significant correlation between metallicity and redshift, {\it or} between metallicity and $E_{\gamma, iso}$ or $E_{\gamma}$. This result is at odds with the previously-predicted and tentatively-observed inverse correlation, and appears to demonstrate that metallicity has no clear impact on the final explosive properties and gamma-ray energy release of an LGRB progenitor.

\section{Spatially-Resolved Host Studies of LGRBs}
\begin{figure}[b]
% \vspace*{-2.0 cm}
\begin{center}
 \includegraphics[width=5in]{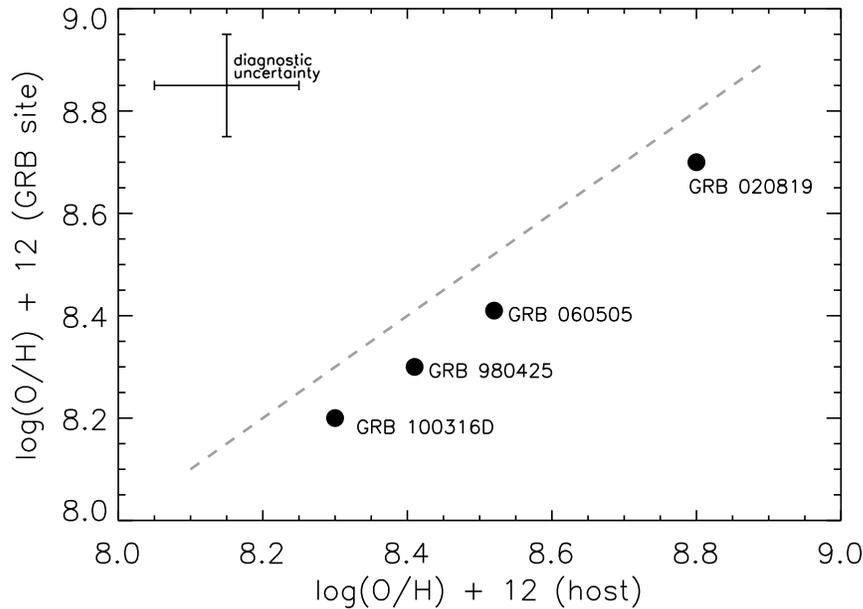} 
% \vspace*{-1.0 cm}
 \caption{Adapted from Levesque et al.\ (2011); explosion site metallicities vs. average host metalicities for the current sample of previously-studied nearby LGRB host galaxies. All four explosion sites fall $\sim$0.1 dex below the theoretical relation where explosion site metallicity and host metallicity are identical, plotted here as a gray dashed line, though this is within the uncertainty of the metallicity diagnostics.}
   \label{fig1}
\end{center}
\end{figure}
It is important to note one strong limitation of current LGRB studies: their reliance on global metallicities. For the majority of LGRB hosts at $z \ge 0.3$, pinpointing the LGRB explosion site and acquiring site-specific spectra within the small, faint host galaxies is a difficult proposition. However, these site-specific studies {\it are} possible for a key sample of seven nearby spatially-resolved LGRB host galaxies. For this subset of hosts we can determine metallicities and star-formation rates directly at the LGRB host site as well as in the surrounding star-forming regions of the galaxy. This allows us to pinpoint the precise environments that produce LGRBs and place these sites in context with their global host galaxy properties. Despite the enormous value of such observations, only a small handful of spatially-resolved LGRB hosts have been previously studied. Christensen et al.\ (2008) obtained integral field unit spectroscopy of the $z = 0.008$ host galaxy of GRB 980425, determining metallicities at 23 different sites across the host. Th\"{o}ne et al.\ (2008) examined spatially-resolved ISM properties in the $z = 0.089$ host galaxy of GRB 060505. Levesque et al.\ (2010d) measured high metallicities at both the nucleus and GRB explosion site within the massive $ z = 0.410$ host galaxy of GRB 020819B.

Most recently, Levesque et al.\ (2011) presented a detailed analysis of the GRB 100316D host environment at $z = 0.059$. By obtaining longslit spectra of the host complex at two different position angles using LDSS3 on Magellan, we were able to extract spatially-resolved profiles for a number of key diagnostic emission features, thus constructing metallicity and star formation rate profiles across the host that focused on both the specific LGRB explosion site and the diffuse emission of the host complex. Based on this analysis, we determined that GRB 100316D happened near the lowest-metallicity and most strong star-forming region of the host complex. However, this work also revealed only a very weak metallicity gradient within the host complex. Combined with the previous studies of nearby LGRB hosts, we found that, within this small sample, ``host" or ``global" metallicities were comparable to metallicities at the GRB explosion sites (Figure 3), suggesting that global metallicities may indeed be valid proxies for explosion site metallicities in higher-redshift LGRB host galaxy studies. Expanding this work to the remaining resolved LGRB hosts (the hosts of GRBs 020903, 030329, and 060218) would allow us to further explore this interesting result.

\section{What's Next?}
Based on the work described above, the role of metallicity in LGRB production and progenitor evolution remains a mystery. LGRBs occur preferentially in low-metallicity environments but do not show any evidence of a cut-off metallicity above which LGRB production is suppressed. There is also no statistically significant correlation between the gamma-ray energy release of LGRBs and the metallicity of their host environments. Finally, it appears that these results cannot be attributed to effects of local metallicities within a globally-sampled host, given that several nearby LGRB hosts show evidence of minimal metallicity gradients and explosion site metallicities that are representative of the global environment. In light of these results, it is worth considering alternative models of LGRBs progenitors and stellar evolution, as well as new analytical means of examining metallicity effects in LGRBs. For example, it is possible that additional comparisons with other explosive properties of LGRBs, such as X-ray fluence or blastwave velocity, could still reveal a correlation with host metallicity.

Alternative progenitor scenarios, such as magnetars or binary channels, could also potentially agree with these observed results. Binary progenitor scenarios in particular are an intriguing possibility. One of the most common binary progenitor scenarios for LGRBs invokes a terminal common envelope phase where the outer envelope is ejected and the stellar cores coalesce. This manner of binary is predicted to occur at a higher rate - but {\it not} exclusively - in low-metallicity environments due to stellar wind effects, with weaker stellar winds permitting the evolution of binaries at closer proximities (Podsiadlowski et al. 2010). A second common progenitor model considers an interim common envelope phase, where the outer envelope is ejected, followed by a contact binary phase. This is also predicted to occur at a higher rate in low-Z environments, due to a widening range of Roche lobe radii that can permit a binary to enter and survive an interim common envelope phase while still maintaining Roche lobe overflow (Linden et al. 2010).

Finally, in addition to new progenitor scenarios, new treatments of stellar evolution with rotation are also compelling. Detailed treatments of differential rotation in massive stars have profound effects on the properties and populations of massive stars (Ekstrom et al. 2012, Levesque et al. 2012), and at low metallicities these effects are expected to be further enhanced (Leitherer 2008). Georgy et al. (2012) recently examined the effects of rotation on the production of evolved massive stars, supernovae, and LGRBs, using the new stellar rotation models of Ekstrom et al.\ (2012) at solar metallicity, and were able to produce favorable conditions for LGRB formation in 40-60M$_{\odot}$ stars at solar metallicity. Indeed, the latest stellar rotation models actually overproduce the predicted rate of LGRBs, although the introduction of additional parameters in the stellar interiors, such as strong coupling of the core to the stellar surface due to interior magnetic fields, could decrease this rate and bring predictions of the models into very good agreement with observations.

Collaborators on this work included Megan Bagley, Edo Berger, Ryan Chornock, Andrew Fruchter, John Graham, Lisa Kewley, and H. Jabran Zahid. The author is supported by NASA through Einstein Postdoctoral Fellowship grant number PF0-110075 awarded by the Chandra X-ray Center, which is operated by the Smithsonian Astrophysical Observatory for NASA under contract NAS8-03060.

\begin{discussion}

\discuss{Nobuyuki}{The GRB host studies have a selection bias against optically dark GRBs. How does this bias affect your conclusion on GRB progenitor models?}

\discuss{Levesque}{It is true that GRB host studies are biased against optically dark GRBs, since with a handful of exceptions it is difficult to confirm their host associations. However, the implications of this work for the future of GRB progenitor modeling remain the same even without this sample. If anything, including this sample of dark GRBs further encourages the pursuit of alternative progenitor scenarios, since some studies suggest that dark GRBs are caused by the production of GRBs in dusty - and potentially higher-metallicity - environments.}

\discuss{Katz}{If restricted to high energies, can a maximum metallicity be ruled out?}

\discuss{Levesque}{Unfortunately, no. The two high-metallicity LGRB hosts in our sample are GRB 020819B and GRB 050826, which have energies on the order of $10^{50}$-$10^{52}$ erg and are consistent with the energies of other ``cosmological" bursts, so there is no apparent maximum metallicity even if we only consider these higher-energy LGRBs.}

\discuss{Vink}{Regarding alternatives for quasi-homogeneous evolution models: we have recently found a subset of rotating Galactic Wolf-Rayet stars from linear polarimetry (Vink et al.\ 2011, A\&A Letters). Their surface velocities are only of order $\sim$100 km s$^{-1}$ but their cores may rotate more rapidly if coupling due to B-fields is not so efficient.}

\discuss{Levesque}{This is a very interesting result, and highlights the importance of modeling stellar rotation and careful treatments of stellar interiors. A very interesting possibility is that LGRB progenitor atmospheres are {\it not} well-coupled to their cores. This would allow high mass loss and angular momentum rates at the stellar surface without removing angular momentum from the core, where a high rotation rate is critical for LGRB production. If LGRB progenitors are sufficiently decoupled in this manner, it would be a possible single-star mechanism for producing LGRBs at high metallicity. Georgy et al.\ (2012) examines this in more detail.}

\discuss{Zhang}{Without looking at the prompt emission properties, what fraction of short GRB hosts can be immediately identified based on the host galaxy information alone?}

\discuss{Levesque}{The nature of host galaxies can, {\it in some cases}, be used to identify whether a burst is short or long. If a GRB is observed in an elliptical galaxy with no star formation, we can safely conclude that it is a ``short" GRB, or a GRB with a compact object progenitor, since LGRBs are restricted to actively star-forming galaxies with young massive star populations. However, the inverse is not true - if a GRB is observed in a star-forming galaxy, we cannot therefore conclude that it is a LGRB, since short GRBs have also been observed in star-forming host galaxies.}

\end{discussion}

\end{document}